\documentclass[preprint,aps,showpacs]{revtex4}
\usepackage{graphicx,epsf}
\usepackage{lscape}
\usepackage{citesort}
\usepackage[colorlinks,citecolor=blue,urlcolor=red,linkcolor=red]{hyperref}
\begin{document}

\title{{\Large{\bf FCNC transition of $B$ to $a_1$ with LCSR }}}

\author{\small
S. Momeni\footnote {e-mail: samira.momeni@phy.iut.ac.ir }, R.
Khosravi  \footnote {e-mail: rezakhosravi @ cc.iut.ac.ir } }

\affiliation{Department of Physics, Isfahan University of
Technology, Isfahan 84156-83111, Iran }

\begin{abstract}
The $B\to a_1 \ell^+ \ell^-$ decays occur by the electroweak penguin
and box diagrams which can be performed through the flavor changing
neutral current (FCNC). We calculate the form factors of the FCNC $B
\to a_1$ transitions in the light--cone sum rules approach, up to
twist--4 distribution amplitudes of the axial vector meson $a_1$.
Forward--backward asymmetry, as well as branching ratios of $B\to
a_1\ell^{+}\ell^{-}$, and $B\to a_1 \gamma$ decays are considered. A
comparison is also made between  our results and the predictions of
other methods.
\end{abstract}


\maketitle

\section{Introduction}
The semileptonic $B$ meson decays are helpful tools for exploring
the Cabibbo, Kabayashi and Maskawa (CKM) matrix elements and CP
violations. These decays usually occur by two various diagrams: 1)
Simple tree diagrams which can be performed via the weak
interaction. 2) Electroweak penguin and  box diagrams which can be
fulfilled through the FCNC transitions in the standard model (SM).
Future study of the FCNC decays can improve our information about:
\begin{itemize}
\item
CP violation, T violation and polarization asymmetries in penguin
diagrams
\item
Exact values for the CKM matrix elements in the weak interactions,
\item
New operators or operators that follow the SM
\item
Development of new physics (NP) and flavor physics beyond the SM.
\end{itemize}
The FCNC decays of $B$ meson are sensitive to NP contributions to
penguin operators. So, to estimate the SM predictions for FCNC
decays and compare these results to the corresponding experimental
values, we can check the SM and search NP.

There is a growing demand for more accurate and reliable
calculations of heavy to light transition form factors in QCD
\cite{Weinzierl,Bagan,BallJ,Zwicky,AKTM,BVLCSR,BallZw_BV,Braun}. The
transition of heavy $B$ meson to light meson $a_1$ is one of the
decays attracted much attention of authors. The form factors of the
transition $B \to a_1 \ell\nu $ have been calculated via such
different approaches as the QCD sum rules (SR) \cite{Aliev}, the
covariant quark model (LFQM) \cite{Chua}, the constituent
quark-meson model (CQM) \cite{Deandrea}, and the Light--cone sum
rules (LCSR) \cite{Kwei,Kwei2}. Also, the $B \to a_1$ decay, as a
FCNC process, has been studied in the perturbative QCD (PQCD)
\cite{Li}, and three-point QCD sum rules (3PSR) \cite{kh}.

In this paper, the FCNC $ B\to a_1 \ell^{+}\ell^{-}$ decays  are
considered with the LCSR.  The LCSR is one of the most effective
tools used to determine non--perturbative parameters of hadronic
states. In this approach, the operator product expansion (OPE) is
performed near the light cone $x^2\approx 0$, while the
non--perturbative hadronic matrix elements are described by the
light cone distribution amplitudes (LCDAs) of increasing twist
instead of the vacuum condensates
\cite{Chernyak,Kolesnichenko,Filyanov,Zhitnitsky,Filyanov2}. The
main purpose of this paper is to calculate the form factors of the
FCNC $B \to a_1$ transitions up to twist--4 distribution amplitudes
of the axial vector meson $a_1$ and to compare the results of these
form factors with those of other approaches.

The paper is organized as follows: In Sec. II, by using the LCSR,
the form factors of $B \to a_1 \ell^+\ell^-$ decays are derived. In
Sec. III, we present the numerical analysis of the LCSR for the form
factors and determine the branching ratio values of the $B\to a_1
\gamma$, and $B\to a_1  \ell^+\ell^-$ decays. Also, the
forward--backward asymmetry of these decays is considered. For a
better analysis, a comparison is made between  our results and the
predictions of other methods.

\section{Transition form factors in the LCSR}
\begin{figure}[th]
\includegraphics[width=10cm,height=3cm]{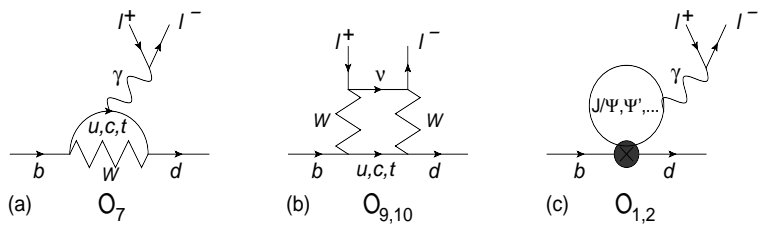}
\caption{(a) and (b) $O_7$ and $O_{9,10}$ short distance
contributions.   (c) $O_{1,2}$ long distance charm-loop
contribution.} \label{F21}
\end{figure}

The $b \to d~ \ell^+ \ell^-$ transition in quark level is explained
by the effective Hamiltonian in the SM as \cite{Munz}:
\begin{equation}\label{eq21}
H_{\rm eff} = - \frac{G_F}{\sqrt{2}} V_{tb}V_{td}^{*}
\sum_{i=1}^{10} C_i(\mu)  O_i(\mu),
\end{equation}
where $V_{tb}$ and $V_{td}$ are the  the CKM matrix elements,
$C_i(\mu)$ and $O_i(\mu)$ are the Wilson coefficients and  the local
operators  respectively, as found in \cite{FGIKL}. The most relevant
contributions to $b \rightarrow d \ell^+ \ell^-$ transitions are: a)
the tree level operators $O_{1,2}$, b) the penguin operator $O_{7}$,
and c) the box operators $O_{9,10}$. The penguin and box operators
are responsible for the short distance (SD) effects in the FCNC $b
\to d$ transition. The current--current operators $O_{1,2}$ involve
an intermediate charm--loop coupled to the lepton pair via the
virtual photon (see Fig. \ref{F21}). This long distance (LD)
contribution has got the same form factor dependence as $C_9$ and
can, therefore, be absorbed into an effective Wilson coefficient
$C^{\rm eff}_9$.
Therefore,  the effective Hamiltonian for $B \rightarrow a_1 \ell^+
\ell^-$  decays which occur via the $b \rightarrow d \ell^+ \ell^-$
loop transition  can be written as:
\begin{eqnarray}\label{eq22}
\cal H_{\rm eff}&=& \frac{G_{F}\alpha}{2\sqrt{2}\pi}
V_{tb}V_{td}^{*}\Bigg[ C_9^{\rm eff}  \overline {d} \gamma_\mu
(1-\gamma_5) b~  \overline {l }\gamma_\mu l + C_{10} \overline {d}
\gamma_\mu (1-\gamma_5) b~ \overline {l } \gamma_\mu \gamma_5 \l
\nonumber\\&-& 2 C_7^{\rm eff}\frac{m_b}{q^2} \overline {d}
~i\sigma_{\mu\nu} q^\nu (1+\gamma_5) b~  \overline {l}  \gamma_\mu l
\Bigg],
\end{eqnarray}
where $C_7^{\rm eff}= C_7-C_5/3-C_6$. The effective Wilson
coefficient, $C^{\rm eff}_9(q^2)$, is given as:
\begin{eqnarray}\label{eq23}
C^{\rm eff}_9 = C_9 + Y_{SD}(q^2)+Y_{LD}(q^2),
\end{eqnarray}
$Y_{SD}(q^2)$ describes the SD contributions from four--quark
operators far away form the $c\bar{c}$ resonance regions, which can
be calculated reliably in perturbative theory. The function
$Y_{LD}(q^2)$ contains the LD contributions coming from the real
$c\bar c$ intermediate states called charmonium resonances. Two
resonances, $J/\psi$ and  $\psi'$, are the narrow. Last four
resonances, $\psi(3370)$, $\psi(4040)$, $\psi(4160)$ and
$\psi(4415)$, are above the $D\bar D$-threshold and as a
consequence, the width is much larger. The explicit expressions of
the $Y_{SD}(q^2)$ and $Y_{LD}(q^2)$ can be found in \cite{Buchalla}
(see also \cite{Munz,LYON, FGIKL}).

To calculate the form factors of the FCNC $B \to a_1$ transition
within the LCSR method, two correlation functions are written as:
\begin{eqnarray}\label{eq24}
\Pi^{V,A}_\mu &=&i \int d^4x e^{iqx} \langle a_1 (p', \varepsilon) |
{\cal{T}} \{\bar{d}(x) \gamma_\mu (1-\gamma_5) b(x)\, j_{B}^{\dag}(0) \}| 0 \rangle,\nonumber\\
\Pi^{T}_{\mu} &=& i \int d^4x e^{iqx} \langle a_1 (p', \varepsilon)
| {\cal{T}} \{\bar{d}(x) \sigma_{\mu\nu} q^{\nu} (1+\gamma_5) b(x)\,
j_{B}^{\dag}(0) \}| 0 \rangle,
\end{eqnarray}
where  $j_B=i \bar{u} \gamma_5 b$  is the interpolating current for
the ${B}$ meson. According to the general philosophy of the LCSR,
the above correlation functions should be calculated in two
different ways. In phenomenological or physical representation, it
is investigated in terms of hadronic parameters. In QCD or
theoretical side, it is obtained in terms of distribution amplitudes
and QCD degrees of freedom. Physical quantities like form factors
are found to equate the coefficient of the same structures from both
representations of the correlation functions through dispersion
relation and apply Borel transformation to suppress the
contributions of the higher states and continuum.

\subsection{Phenomenological side}

By considering phenomenological representation, a complete set of
hadrons with the same quantum numbers as the interpolating current
operator $j_B$ is inserted in the correlation functions. After
isolating the pole mass term of the $B$ meson and  applying Fourier
transformation as well as the dispersion relation, we obtain:
\begin{eqnarray}\label{eq25}
\Pi^{V,A}_{\mu} (p',p)&=& \frac{\langle a_{1}(p',\varepsilon)|\bar
{d}\,\gamma_{\mu}(1-\gamma_5)\,b|B(p)\rangle \langle
B(p)|\bar{b}~ i \gamma_5\,u|0\rangle}{m^2_B-p^2}+\frac{1}{\pi}\int_{s_0}^{\infty}  \frac{ \rho^{h(V,A)}_{\mu}(s) }{s-p^2}ds,\nonumber \\
\Pi^{T}_{\mu}(p',p) &=& \frac{\langle a_{1}(p',\varepsilon)|\bar
{d}\,\sigma_{\mu\nu}q^{\nu}(1+\gamma_5)\,b|B(p)\rangle \langle
B(p)|\bar{b}i \gamma_5\,u
|0\rangle}{m^2_B-p^2}+\frac{1}{\pi}\int_{s_0}^{\infty}  \frac{
\rho^{h(T)}_{\mu}(s) }{s-p^2}ds,
\end{eqnarray}
where $\rho^h_{\mu}$ shows the spectral density of the higher
resonances and the continuum states in the hadronic representation.
These spectral densities are approximated by evoking the
quark--hadron duality assumption,
\begin{eqnarray}\label{eq26}
\rho^h_{\mu}(s)&=&\rho^{QCD}_{\mu}(s)\theta(s-s_0),
\end{eqnarray}
where $\rho^{QCD}_{\mu}(s)$ is the perturbative  QCD spectral
density investigated from the theoretical side of the correlation
function. The threshold $s_0$ is chosen near the squared mass of the
lowest $B$ meson state.

The matrix elements $\langle a_{1}(p',\varepsilon) | \bar{d}
\gamma_\mu (1-\gamma_{5}) b | B(p)\rangle$ and $\langle
a_{1}(p',\varepsilon) | \bar{d} \sigma_{\mu\nu} q^\nu (1+\gamma_5)b
| B(p) \rangle$ are parameterized in terms of the form factors as
follows:
\begin{eqnarray}\label{eq27}
\langle a_{1}(p',\varepsilon) |\bar{d}\gamma_\mu (1-\gamma_{5})b|
B(p) \rangle &=&i  {2 A (q^2) \over m_{B}- m_{a_{1}}}
\epsilon_{\mu\nu\alpha\beta}\, \varepsilon^{*\nu} p^\alpha p'^\beta
- V_{1}(q^2)  \varepsilon_\mu^{*}(m_{B} - m_{a_{1}})  \nonumber
\\&-&
\frac{V_{2}(q^2)}{m_B-m_{a_{1}}} (\varepsilon^{*}. q) (p+p')_\mu + 2
m_{a_{1}} \frac{( \varepsilon^{*}.q)}{q^2} q_\mu
[V_{3}(q^2) - V_{0}(q^2)], \nonumber\\
\langle a_{1}(p',\varepsilon) |\bar{d}\sigma_{\mu\nu}
q^\nu(1+\gamma_5)b| B(p) \rangle &=& 2 T_{1} (q^2)
\epsilon_{\mu\nu\alpha\beta} \varepsilon^{*\nu} p^\alpha p'^\beta +
iT_{2} (q^2) \Big[(m_{B}^2 - m_{a_1}^2) \varepsilon^{*}_\mu -
(\varepsilon^{*}.q)
(p+p')_\mu \Big] \nonumber \\
&-&iT_3 (q^2) (\varepsilon^{*}. q)\Bigg[ q_\mu -\frac{ q^2}{{m_{B}^2
- m_{a_1}^2}} (p+p')_\mu \Bigg],
\end{eqnarray}
where $q = p-p'$ is the momentum transfer of the $Z$  boson
(photon), and $\varepsilon^{*\nu}$ is the polarization vector of the
axial vector meson $a_1$. It should be noted that $V_{0}(0)=V_{3}(0)
$. On the other hand, the identity
$\sigma_{\mu\nu}\gamma_{5}=-\frac{i}{2}\epsilon_{\mu\nu\alpha\beta}\sigma^{\alpha\beta}$
implies that $T_{1}(0)=T_{2}(0) $\cite{Colangelo}. Also, $V_{3} $
can be written as a linear combination of  $V_{1}  $ and $V_{2} $:
\begin{eqnarray}\label{eq28}
V_{3}(q^{2})=\frac{m_{B}-m_{a_1}}{2m_{a_1}}
V_{1}(q^{2})-\frac{m_{B}+m_{a_1}}{2m_{a_1}} V_{2}(q^{2}).
\end{eqnarray}

Taking into account the second matrix element in Eq. (\ref{eq25}) as
$ \langle B(p_{B})|\bar{b}i \gamma_5\,u\,|0\rangle=\frac{f_{B}
m_{B}^{2} }{m_b}$, where $f_{B}$ is the $B$ meson decay constant and
$m_b$ is the $b$ quark mass, we can obtain these hadronic
representations for $\Pi_{\mu}^{A,V}$ and $\Pi_{\mu}^{T}$ as:
\begin{eqnarray}\label{eq29}
\Pi_\mu^{A,V} &=& -\frac{f_B m_B^2}{m_b} \frac{1}{p^2 - m_B^2}
\Bigg\{i \frac{2 A(q^2)}  {m_B - m_{a_1}}
\epsilon_{\mu\nu\alpha\beta} \varepsilon^{*\nu} p^\alpha p'^\beta
-V_1(q^2) \varepsilon_\mu^{*}(m_B-m_{a_1}) \nonumber \\
&-& \frac{V_2(q^2)}{ m_B-m_{a_1}}(\varepsilon^{*}. q) {(p+p')}_\mu +
2 m_{a_1} \frac{(\varepsilon^{*}.q)}{q^2} q_{\mu}
[V_3(q^2)-V_0(q^2)] \Bigg\}+\frac{1}{\pi}\int_{s_0}^{\infty} \frac{
\rho_{\mu}^{h(A,V)}(s) }{s-p^2}ds,\nonumber\\
\Pi_\mu^{T} &=& -\frac{f_B m_B^2}{m_b} \frac{1}{p^2 - m_B^2}
\Bigg\{2 T_{1} (q^2) \epsilon_{\mu\nu\alpha\beta} \varepsilon^{*\nu}
p^\alpha p'^\beta + iT_{2} (q^2) \Big[(m_{B}^2 - m_{a_1}^2)
\varepsilon^{*}_\mu - (\varepsilon^{*}.q)
(p+p')_\mu \Big] \nonumber \\
&-&iT_3 (q^2) (\varepsilon^{*}. q)\Bigg[ q_\mu -\frac{ q^2}{{m_{B}^2
- m_{a_1}^2}} (p+p')_\mu \Bigg]
\Bigg\}+\frac{1}{\pi}\int_{s_0}^{\infty} \frac{ \rho_{\mu}^{h(T)}(s)
}{s-p^2}ds.
\end{eqnarray}

\subsection{Theoretical side}
Now, the QCD or the theoretical part of  the correlation functions
should be calculated. The calculation for the  defined  correlators
in the region of large space--like momenta  is based on the
expansion of the ${\cal T}$-product of the currents  near the light
cone $x^2 = 0$. After contracting $b$ quark field, we get
\begin{eqnarray}\label{eq31}
\Pi^{A,V}_{\mu} &=& \int d^4x e^{iqx} \langle a_1 (p', \varepsilon)
|
\bar{d}(x) \gamma_\mu (1-\gamma_5) S^{b}(x,0) \gamma_5\,u(0) | 0 \rangle, \nonumber \\
\Pi^{T}_{\mu} &=&  \int d^4x e^{iqx} \langle a_1 (p', \varepsilon) |
\bar{d}(x) \sigma_{\mu\nu}\,q^{\nu} (1+\gamma_5) S^{b}(x,0)
\gamma_5\,u(0) | 0\rangle,
\end{eqnarray}
where  $S^{b}(x,0)$ is the full propagator of the $b$ quark in
presence of the background gluon field as:
\begin{eqnarray}\label{eq32}
\,S^{b}(x)&=& \int \frac{d^4k}{(2\pi)^4} e^{-ikx} \frac{\not\!k +
m_b}{k^2-m_b^2}-g_s\int
\frac{d^4k}{(2\pi)^4}e^{-ikx}\int_0^1du\left[\frac{1}{2}\frac{k\!\!\!/+m_b}{(m_b^2-k^2)^2}G_{\mu\nu}(ux)\sigma^{\mu\nu}
\right.\nonumber\\
&+&\left.\frac{1}{m_b^2-k^2}ux_\mu G^{\mu\nu}(ux)\gamma_\nu\right],
\end{eqnarray}
where $G_{\mu\nu}$ is the gluon field strength tensor and  $g_s$ is
the strong coupling constant. In the present work, contributions
with two gluons as well as four quark operators are neglected
because their contributions are small. Using Fierz rearrangement
formula, Eq. (\ref{eq31}) can be rewritten  as:
\begin{eqnarray}\label{eq33}
\Pi_\mu^{A,V}&=&-\frac{i}{4} \int d^4x e^{iqx} \Big[\mbox{\rm Tr}
\{\gamma_\mu (1-\gamma_5) S^{b}(x)\,\gamma_5 \Gamma_{\alpha}\}
\Big]\langle a_{1} \vert
\bar{d}(x)\, \Gamma^{\alpha} u(0) \vert 0 \rangle,\nonumber\\
\Pi_{\mu}^{T}&=&-\frac{i}{4} \int d^4x e^{iqx} \Big[\mbox{\rm Tr}
\{\sigma_{\mu\nu} (1+\gamma_5) S^{b}(x)\,\gamma_5 \Gamma_{\alpha}\}
\Big]\,q^{\nu}\,\langle a_{1} \vert \bar{d}(x)\, \Gamma^{\alpha}
u(0) \vert 0 \rangle,
\end{eqnarray}
where $\Gamma_{\alpha}$ is the full set of the Dirac matrices,
$\Gamma_{\alpha}= (I,~\gamma_5,~\gamma_\mu,~\gamma_{\mu}
\gamma_5,~\sigma_{\mu\nu})$. In order to calculate the theoretical
part of the correlator functions in Eq. (\ref{eq33}), the matrix
elements of the nonlocal operators between $a_{1}$ meson and vacuum
states are needed. Two--particle distribution amplitude up to
twist--4 for the axial vector meson $a_{1}$ is given in \cite{Kwei}:
\begin{eqnarray}\label{eq34}
\langle a_1(p',\varepsilon) | \bar{d}_\alpha (x) u_\delta (0) | 0
\rangle &=& -\frac{i}{4}  \int_0^1 du~ e^{i u  p'. x}\Bigg\{ f_{a_1}
m_{a_1} \Bigg[ \not\! p'\gamma_5 \frac{\varepsilon^*. x}{p'.x}
\Phi_\parallel(u) +\Bigg( \not\! \varepsilon^* -\not\! p'
\frac{\varepsilon^*. x}{p'.x}\Bigg)\gamma_5 g_\perp^{(a)}(u) \nonumber\\
&-& \not\! x\gamma_5 \frac{\varepsilon^*. x}{2(p'.x)^2} m_{a_1}^2
\bar g_3(u) + \epsilon_{\mu\nu\rho\sigma} \varepsilon^{*\nu}
p'^{\rho} x^\sigma \gamma^\mu
\frac{g_\perp^{(v)}(u)}{4}\Bigg] \nonumber\\
&+& \,f^{\perp}_{a_1} \Bigg[ \frac{1}{2}( \not\! p'\not\!\epsilon^*-
\not\!\epsilon^* \not\! p' ) \gamma_5\,\Phi_\perp(u) - \frac{1}{2}(
\not\! p'\not\! x- \not\! x \not\! p' ) \gamma_5 \frac{\epsilon^*.
x}{(p'.x)^2} m_{a_1}^2 \bar
h_\parallel^{(t)} (u) \nonumber\\
&+& i \Big(\epsilon^*. x\Big) m_{a_1}^2 \gamma_5
\frac{h^{(p)}_\parallel (u)}{2} \Bigg]\Bigg\}_{\delta\alpha},
\end{eqnarray}
where for $x^{2}\neq 0$, we have
\begin{eqnarray*}\label{eq35}
\bar g_3 (u) &=& g_3(u) +\Phi_\parallel -2 g_\perp^{(a)}(u),\nonumber\\
\bar h_\parallel^{(t)} &=& h_\parallel^{(t)}- \frac{1}{2}
\Phi_\perp(u).
\end{eqnarray*}
In Eq. (\ref{eq34}), $\Phi_\parallel$, $\Phi_\perp$ are the twist-2,
$g_\perp^{(a)}$, $g_\perp^{(v)}$, $h_\parallel^{(t)}$ and
$h_\parallel^{(p)}$ are twist-3, and $g_3$ is twist-4 functions. The
definitions for $\Phi_\parallel$, $\Phi_\perp$, $g_\perp^{(a)}$,
$g_\perp^{(v)}$, $h_\parallel^{(t)}$, $h_\parallel^{(p)}$ and $g_3$
are given in Appendix \ref{app:fun-def}.

Two--particle chiral--even distribution amplitudes are given by
\cite{Kwei}:
\begin{eqnarray}\label{eq36}
\langle a_1(p',\varepsilon)|\bar d(x) \gamma_\mu \gamma_5
u(0)|0\rangle &=& i f_{a_1} m_{a_1}\int_0^1 du \,  e^{i u p'. x}
\Bigg\{ p'_\mu \frac{\varepsilon^{*}. x}{p'. x} \Phi_\parallel(u)
+\left( \varepsilon_{\mu}^{*} -p'_\mu \frac{\varepsilon^{*}.
x}{p'.x}\right) g_\perp^{(a)}(u) \nonumber\\& -& \frac{1}{2}x_{\mu}
\frac{\varepsilon^{*}. x }{(p'. x)^{2}} m_{a_1}^{2} \bar g_{3}(u)
+{\cal O}(x^2) \Bigg\}~,\nonumber\\
\langle a_1 (p',\varepsilon)|\bar d(x) \gamma_\mu u(0)|0\rangle & =
& - i f_{a_1} m_{a_1} \epsilon_{\mu\nu\rho\sigma} \varepsilon^{*\nu}
p'^{\rho} x^\sigma \int_0^1 du \, e^{i u \, p'. x}\Bigg\{
\frac{g_\perp^{(v)}(u)}{4}+{\cal O}(x^2)\Bigg\},
\end{eqnarray}
also, two--particle chiral--odd distribution amplitudes are defined
by:
\begin{eqnarray}\label{eq37}
\langle a_1(p',\varepsilon)|\bar d(x) \sigma_{\mu\nu}\gamma_5 u(0)
|0\rangle & =&  f_{a_1}^{\perp} \int_0^1 du \, e^{i u p'. x}
\Bigg\{(\varepsilon^{*}_{\mu} p'_{\nu} - \varepsilon_{\nu}^{*}
p'_{\mu}) \Phi_\perp(u) + \frac{m_{a_1}^2\,\varepsilon^{*}. x}{(p'.
x)^2}(p'_\mu x_\nu - p'_\nu  x_\mu) \bar{h}_\parallel^{(t)}
\nonumber\\&+&{\cal O}(x^2)\Bigg\}, \nonumber \\
\langle a_1(p',\varepsilon)|\bar d(x) \gamma_5 u(0) |0\rangle &=&
f_{a_1}^\perp m_{a_1}^2 (\varepsilon^{*}. x)\int_0^1 du \, e^{i u
p'. x}\Bigg\{\frac{h_\parallel^{(p)}(u)}{2}+ {\cal O}(x^2)\Bigg\}.
\end{eqnarray}
In these expressions, $f_{a_1}$ and $f_{a_1}^{\perp}$ are decay
constants of the axial vector meson $a_1$ defined as:
\begin{eqnarray}\label{eq38}
\langle a_{1}(p',\varepsilon)|\bar d(0) \gamma_\mu \gamma_5
u(0)|0\rangle
&=& if_{a_1}\, m_{a_1} \epsilon^{*}_\mu,\nonumber\\
\langle a_{1}(p',\varepsilon)| \bar d(0) \sigma_{\mu\nu}\gamma_5
u(0)|0\rangle & = & f_{a_1}^{\perp} a_0^{\perp} \,
(\epsilon^{*}_{\mu} p'_{\nu} - \epsilon_{\nu}^{*} p'_{\mu}),
\end{eqnarray}
where $a_0^{\perp}$ refers to the zeroth Gegenbauer moments of
$\Phi_\perp$. It should be noted that $f_{a_1}$ is
scale--independent and conserves   $G$-parity, but $f_{a_1}^{\perp}$
is scale--dependent and violates $G$-parity.

\begin{figure}[th]
\includegraphics[width=10cm,height=3cm]{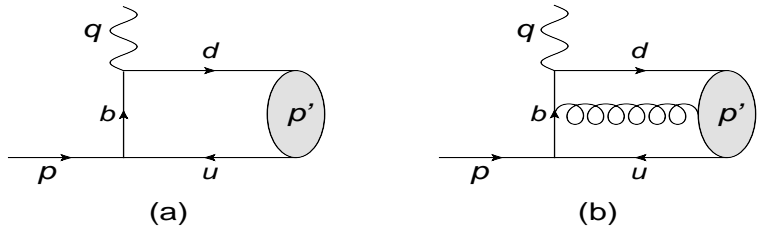}
\caption{Leading--order terms in the correlation functions involving
the two--particle (a) and three--particle (b).} \label{F1}
\end{figure}

Three--particle distribution amplitudes are defined  as:
\begin{eqnarray}\label{eq39}
\langle a_1(p',\varepsilon) | \bar{d}(x) \gamma_\alpha \gamma_5 g_s
G_{\mu\nu} (ux) u (0) | 0 \rangle &=& p'_\alpha (p'_\nu
\varepsilon^{*}_\mu - p'_\mu \varepsilon^{*}_\nu) f_{3, a_1}^A {\cal A} +\cdots,  \nonumber\\
\langle a_1(p',\varepsilon) | \bar{d}(x) \gamma_\alpha g_s
\widetilde{G}_{\mu\nu} (ux) u (0) | 0 \rangle&= & i p'_\alpha
(p'_\mu \varepsilon^{*}_\nu - p'_\nu \varepsilon^{*}_\mu) f_{3,
a_1}^V {\cal V} +\cdots,
\end{eqnarray}
where
$\widetilde{G}_{\mu\nu}=\frac{1}{2}\epsilon_{\mu\nu\rho\lambda}G^{\rho\lambda}$.
The value of coupling constants $f_{3, a_1}^V $ and $f_{3, a_1}^A$
for $a_1$ meson at $\mu=1\rm{GeV}$ is: $f_{3, a_1}^V=(0.0055 \pm
0.0027)\, \rm{GeV^2}$ and $f_{3, a_1}^A=(0.0022 \pm 0.0009)\,
\rm{GeV^2}$  \cite{Kwei2}. The three--parton chiral--even
distribution amplitudes  $\cal A$ and $\cal V$ in Eq. (\ref{eq39})
are defined as:
\begin{eqnarray}\label{eq310}
{\cal A}&=&  \int {\cal D}\underline{\alpha}
\,e^{ip'.x(\alpha_{1}+u\alpha_{3})}{\cal A} (\alpha_{i}),\nonumber\\
{\cal V}&=&  \int {\cal D}\underline{\alpha}
\,e^{ip'.x(\alpha_{1}+u\alpha_{3})}{\cal V} (\alpha_{i}),
\end{eqnarray}
where ${\cal A}(\alpha_{i})$ and ${\cal V}(\alpha_{i})$ can be
approximately written as \cite{Kwei2}:
\begin{eqnarray}\label{eq311}
{\cal A}(\alpha_{i})&=&5040
(\alpha_{1}-\alpha_{2})\alpha_{1}\alpha_{2}\alpha_{3}^2
+360\alpha_{1}\alpha_{2}\alpha_{3}^2 \Big[ \lambda^A_{a_1}+
\frac{\sigma^A_{a_1}}{2}(7\alpha_3-3)\Big],
\nonumber\\
{\cal V}(\alpha_{i})&=&360\alpha_{1}\alpha_{2}\alpha_{3}^2 \Big[ 1+
\frac{\omega^V_{a_1}}{2}(7\alpha_3-3)\Big] +5040
(\alpha_{1}-\alpha_{2})\alpha_{1}\alpha_{2}\alpha_{3}^2
\sigma^V_{a_1},
\end{eqnarray}
where  $\alpha_{1}$, $\alpha_{2}$, and $\alpha_3$ are the momentum
fractions carried by $d$, $\bar u$ quarks and gluon,  respectively,
in the axial vector meson $a_1$. The integration measure is defined
as:
\begin{equation}\label{eq312}
\int {\cal D}\underline{\alpha} \equiv \int_0^1 d\alpha_{1} \int_0^1
d\alpha_{2}\int_0^1 d\alpha_3 \,\delta(1-\sum \alpha_i).
\end{equation}
Diagrammatically, the contributions of two-- and three--particle
LCDAs to the correlation functions are depicted in Fig. \ref{F1}.

In this step, inserting  the full propagator (Eq. (\ref{eq32})) and
two--particle as well as three--particle LCDAs (Eqs.
(\ref{eq34}-\ref{eq39})) in the correlation functions (Eq.
(\ref{eq33})), traces and then integrals should be calculated. To
estimate these calculations, we have used identities as:
\begin{eqnarray}\label{eq314}
\epsilon_{\alpha\beta\gamma\sigma}\,\epsilon^{\alpha}_{\mu\nu\lambda}&=&(\delta_{\nu\beta}\,\delta_{\lambda\sigma}\,\delta_{\mu\gamma}
-\delta_{\nu\beta}\,\delta_{\lambda\gamma}\,\delta_{\mu\sigma}+\delta_{\lambda\beta}\,\delta_{\nu\gamma}\,\delta_{\mu\sigma}
-\delta_{\lambda\beta}\,\delta_{\mu\gamma}\,\delta_{\nu\sigma}
+\delta_{\mu\beta}\,\delta_{\lambda\gamma}\,\delta_{\nu\sigma}
-\delta_{\mu\beta}\,\delta_{\nu\gamma}\,\delta_{\lambda\sigma})\nonumber\\
\epsilon_{\sigma\alpha\beta\mu}\,\epsilon^{\alpha\beta\rho\lambda}&=&
2\,(\delta^{\rho}_{\sigma}\,\delta^{\lambda}_{\mu}-\delta^{\lambda}_{\sigma}\,\delta^{\rho}_{\mu}).
\end{eqnarray}

Now, to get the LCSR for the calculations of the $B \to a_1$ form
factors, we equate the coefficients of the corresponding structures
from both phenomenological and theoretical sides of the correlation
functions and apply Borel transform with respect to the variable $p$
as:
\begin{eqnarray}\label{eq316}
B_{p^2}(M^2)\frac{1}{\left(
p^{2}-m_{B}^{2}\right)^{n}}&=&\frac{(-1)^{n}}{\Gamma(n)}\frac{e^{-\frac{m_{B}^{2}}{M^{2}}}}{(M^{2})^{n}},
\end{eqnarray}
in order to suppress the contributions of the higher states and
continuum as well as eliminate the subtraction terms. Thus, the form
factors are obtained via the LCSR. The explicit expressions for the
form factors are presented in Appendix \ref{app:form factors}.

\section{Numerical analysis}
In this section, we present our numerical analysis for the form
factors and branching ratios of the $B\to a_1 \ell^+\ell^-$ decays.
In this work, masses  are taken in $\mbox{GeV}$ as $m_b=4.81\pm
0.03$, $m_{\mu}=0.11$, $m_{\tau}=1.77$, $m_{a_1}=1.23\pm 0.04$, and
$m_{B}=5.27\pm0.01$ \cite{pdg}. The  $f_{B}=(0.19 \pm 0.02)
~\mbox{GeV}$  is in agreement with the QCD sum rule result with
radiative corrections \cite{Kwei}. The $G$-parity violating decay
constant for $a_1$ meson is defined by $f^\perp_{a_1}$ and is equal
to $f_{a_1}=(0.23\pm0.01)\,\rm{GeV}$ at the energy scale
$\mu=1\,\rm{GeV}$ \cite{Kwei}. The suitable threshold parameter
$s_0$ is chosen as $s_0=(33 \pm 1)~\mbox{GeV}^2$, which corresponds
to the sum rule calculation \cite{BallZw_BV}. Also, we need to know
Gegenbauer moments of $\Phi_\perp$, $\Phi_\parallel$, and $G$-parity
conserving parameters of three--parton LCDAs for $a_1$ meson at the
scale $\mu=1\,\rm{GeV}$ given in Table \ref{TR1}. It
should be noted that the value of other parameters such as
$\sigma^{A}_{a_1},\, \sigma^{V}_{a_1},\, \sigma^{\perp}_{a_1},\,
\lambda_{a_1}^{A},\, \zeta^V_{3, a_1},\,a_{0}^{\parallel},\,
a_{1}^{\parallel}$ and $ a_{2}^{\perp}$ is zero for meson $a_1$ \cite{Kwei2}.
\begin{table}[th]
\caption{The Gegenbauer moments of $\Phi_\perp$ and $\Phi_\parallel$
for $a_1$ meson and  twist--3  LCDAs parametrs at $\mu=1\rm{GeV}$.}
\label{TR1}
\begin{ruledtabular}
\begin{tabular}{cccccc}
LCDAs parametrs &$a_1^{\perp}$&$a_2^{\parallel}$&$\zeta^\perp_{3,
a_1}$&$\omega^{\perp}_{a_1}$&$\omega^{V}_{a_1}$ \\
\hline \rm{Value} & $-1.04\pm 0.34 $ & $-0.02\pm 0.02$ &
$-0.009\pm0.001$
& $-3.70 \pm 0.40$& $-2.90 \pm 0.90$\\
\end{tabular}
\end{ruledtabular}
\end{table}

We should obtain the region for the Borel mass parameter so that our
results for the form factors of the $B\to a_1$ decays would be
almost insensitive to variation of $M^2$. We find that the
dependence of the form factors on $M^2$ is small in the interval
$M^2\in[6, 10]~\rm GeV^2$.

Using all these input values and parameters, we can present form
factor values at the zero transferred momentum square $q^2=0$ in
Table \ref{TR2}. The errors in Table \ref{TR2} are estimated by the
variation of the Borel parameter $M^2$, the variation of the
continuum threshold $s_0$, the variation of $b$ quark mass, and the
parameters of the LCDAs. The main uncertainty comes from LCDAs
$\Phi_\perp(u)$ and $b$ quark mass $m_b$, while the other
uncertainties are small, constituting a few percent.
\begin{table}[th]
\caption{The $B \to a_1$  form-factors at zero momentum transfer.
} \label{TR2}
\begin{ruledtabular}
\begin{tabular}{ccccccccc}
\rm{Form factors} &${A}(0)$&${V}_{1}(0)$&${V}_{2}(0)$&${V}_{0}(0)$&$T_{1}(0)= T_{2}(0)$&$T_{3}(0)$ \\
\hline
\rm{Value} & ${0.42}{\pm 0.16} $ & ${0.68}{\pm
0.13} $ & ${0.31}{\pm 0.16} $
& ${0.30}{\pm 0.18} $ & ${0.44}{\pm 0.28} $ & ${0.41}{\pm 0.18} $\\
\end{tabular}
\end{ruledtabular}
\end{table}

The $b$ quark propagator in Eq. (\ref{eq32}) consists of the free
propagator as well as the one--gluon term. Considering only the free
propagator in the QCD calculations, the value of $V_{1}$ at the zero
transferred momentum square $q^2=0$ is $0.67$, which is about $\sim
99 \% $ of the total value, while the contribution of the other part
of the propagator is  about $1 \%$. Table \ref{TR3} shows the
contribution of the $b$ quark free propagator in the form factor
calculations at $q^2=0$.
\begin{table}[th]
\caption{Contribution of the  $b$ quark free propagator in the form
factor values at $q^2=0$.} \label{TR3}
\begin{ruledtabular}
\begin{tabular}{ccccccccc}
Propagator &${A}(0)$&${V}_{1}(0)$&${V}_{2}(0)$&${V}_{0}(0)$&$T_{1}(0) $&$ T_{2}(0)$&$T_{3}(0)$ \\
\hline Free propagator& $0.41$ & $0.67 $ & $0.30 $ & $0.29 $ & $
0.43$ & $ 0.42$ & $0.39 $
\end{tabular}
\end{ruledtabular}
\end{table}
As can be seen in Table \ref{TR3}, the main contribution comes from
the free propagator. So by taking into account the full propagator  instead of the
free  propagator, correction made in the form factor values at the
zero transferred momentum square $q^2=0$ is very small.

\begin{figure}
\includegraphics[width=6cm,height=5cm]{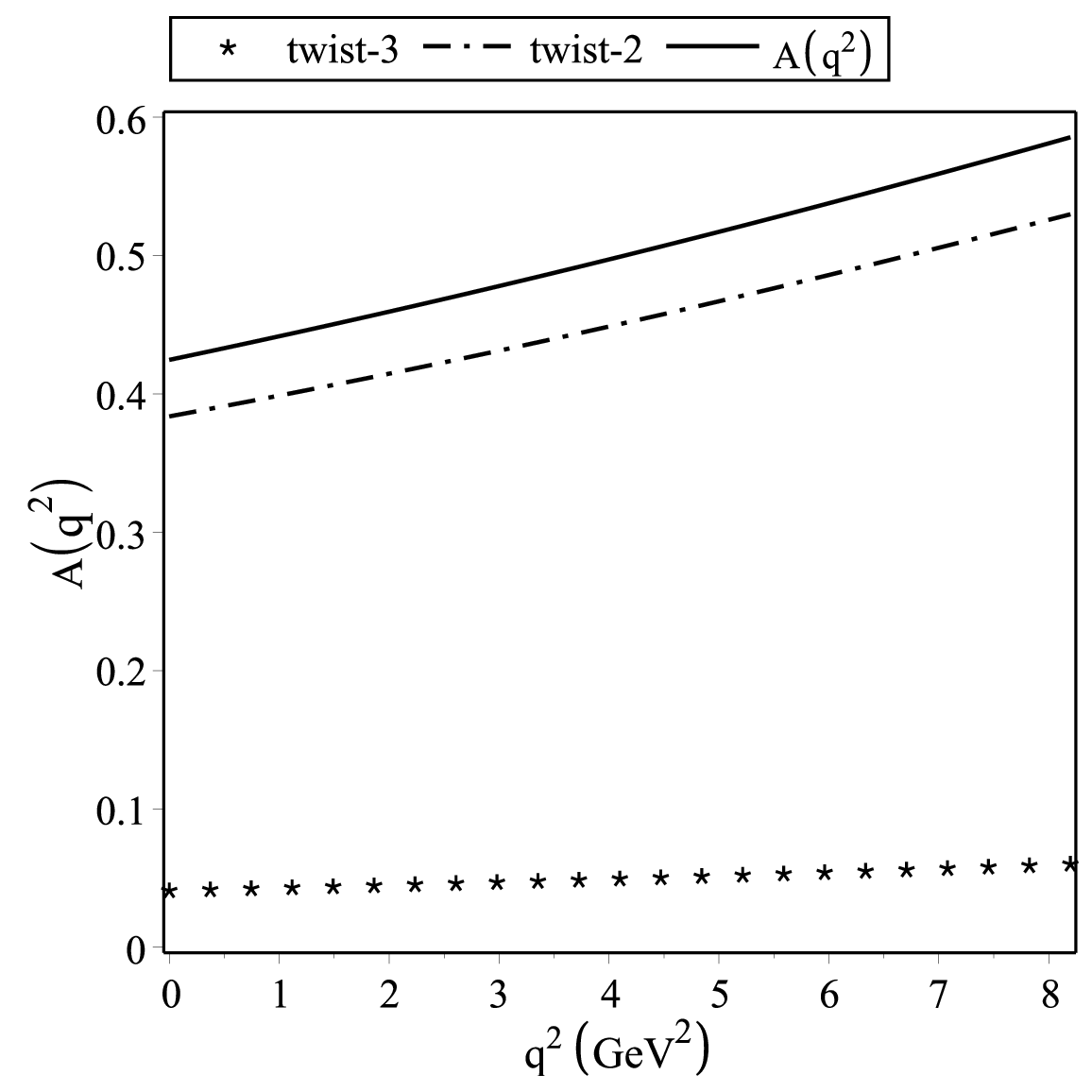}
\caption{Form factor $A$ on $q^2$ as well as the contributions of
twist--2 and twist--3 DAs in this form factor.}\label{FR1}
\end{figure}

In this work, the form factors are estimated in the LCSR approach up
to twist--4 distribution amplitudes of the axial vector meson $a_1$.
Our calculations show that the most contribution comes from twist--2
functions for all form factors. Also, the LCDAs $\Phi_{\perp}$ plays
the most important role in this contribution. Fig. \ref{FR1} depicts
the twist--2 and twist--3 contributions in the form factor formula
$A(q^2)$. In this form factor, the twist--4 function does not
contribute.
Several authors have calculated the form factors of the $B\to a_{1}
\ell^+ \ell^-$ decay via  different approaches. To compare the
different results, we should rescale them according to the form
factor definition in Eq. (\ref{eq23}). Table \ref{T2} shows the
values of the rescaled  form factors at $q^2=0$ according to
different approaches.
{\begin{table}[th]
\caption{Transition form factors of the $B\to a_1 \ell^+ \ell^-$ at
$q^2=0$ in various theoretical approaches. The results of other
methods have been rescaled according to the form factor definition
in Eq. (\ref{eq23}).}
 \label{T2}
\begin{ruledtabular}
\begin{tabular}{ccccccc}
Theoretical approaches&${A}(0)$&${V}_{1}(0)$&${V}_{2}(0)$&${V}_{0}(0)$&$T_{1}(0)=T_{2}(0)$&$T_{3}(0)$ \\
\hline
CQM\cite{Li} &$0.26$& $0.43$ &$0.14$ & $0.34$&$0.34$& $0.19$ \\
3PSR \cite{kh} &$0.31$&$0.52$&  $0.25$ & $0.76$&$0.37$&$0.41$  \\
This Work&$0.42$&$0.68$&$0.31$&$0.30$&$0.44$&$0.41$
\end{tabular}
\end{ruledtabular}
\end{table}
In order to extend our results  to  the whole physical region $4
m_\ell^2 \le q^2 \le (m_B-m_{a_1})^2$, we use the following
parametrization of the form factors with respect to $q^2$ as:
\begin{eqnarray}\label{eq41}
F_{k}(q^{2})=\frac{F_{k}(0)}{1-\alpha\,s+\beta\,s^2},
\end{eqnarray}
where $s= q^2 / m_B^2$ and $F_{k}(q^{2})$ denote for the form
factors, $A$, $V_{i}\,(i=0, 1, 2)$ and $T_{j}\, (j=1, 2, 3)$. The
values of $F_{k}(0)$, $\alpha$ and $\beta$ for  the parameterized
form factors are given in Table \ref{T1}.
\begin{table}[th]
\caption{The parameter values for the fitted form factors.}
\label{T1}
\begin{ruledtabular}
\begin{tabular}{cccccccc}
$\mbox{Form factor}$& ${A}(0)$&${V}_{1}(0)$&${V}_{2}(0)$&${V}_{0}(0)$&$T_{1}(0)$&$T_{2}(0)$&$T_{3}(0)$ \\
\hline
$F(0)$& $0.42$&$0.68$&$0.31$&$0.30$&$0.44$&$0.44$&$0.41$\\
$\alpha$& $1.09$&$0.73$&$0.84$&$0.77$&$0.57$&$0.63$&$0.40$\\
$\beta$& $0.55$&$0.35$&$0.47$&$0.37$&$0.38$&$0.32$&$3.58$
\end{tabular}
\end{ruledtabular}
\end{table}
The fitted form factors with respect to $q^2$ are shown in Fig.
\ref{F2}.
\begin{figure}
\includegraphics[width=6cm,height=5cm]{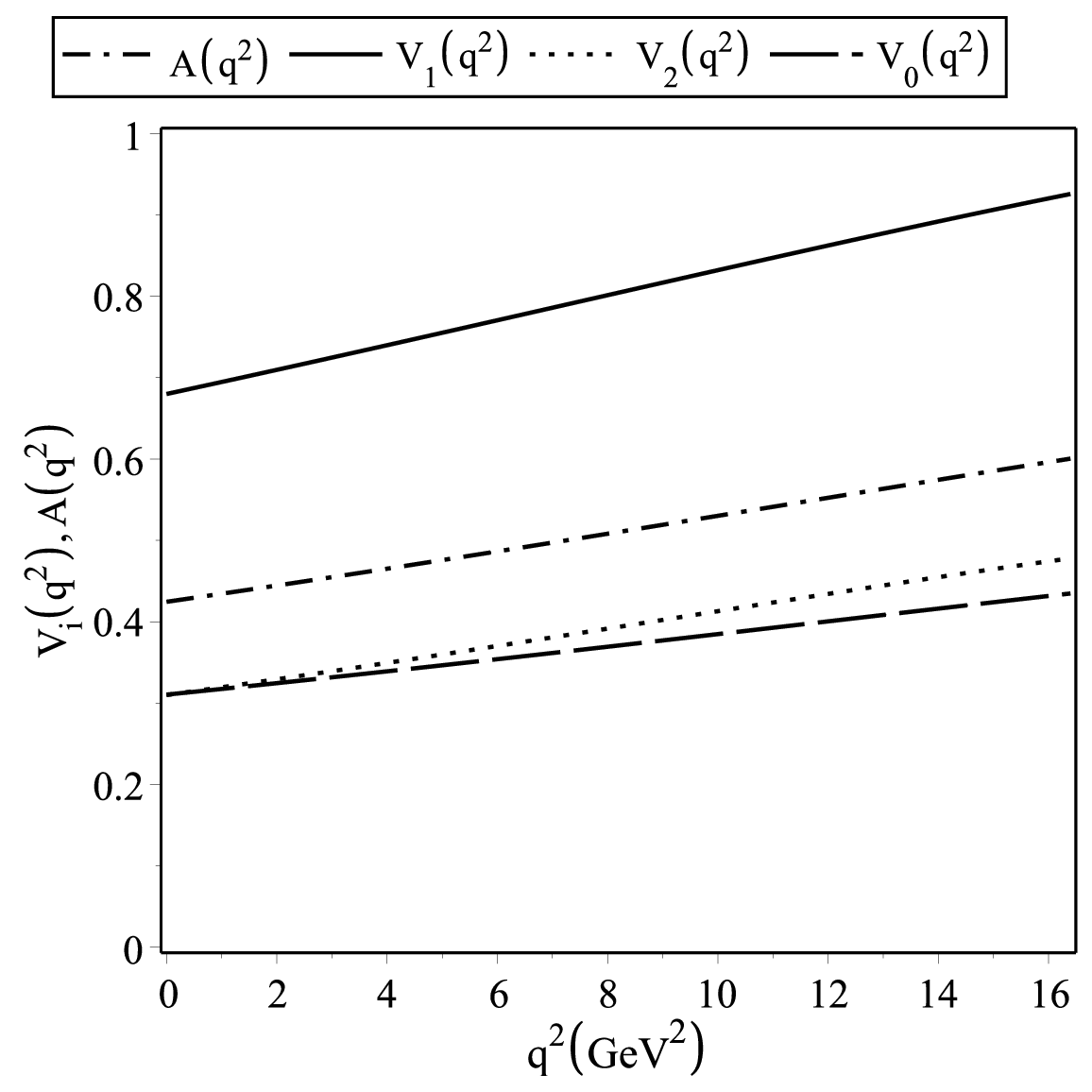}
\includegraphics[width=6cm,height=5cm]{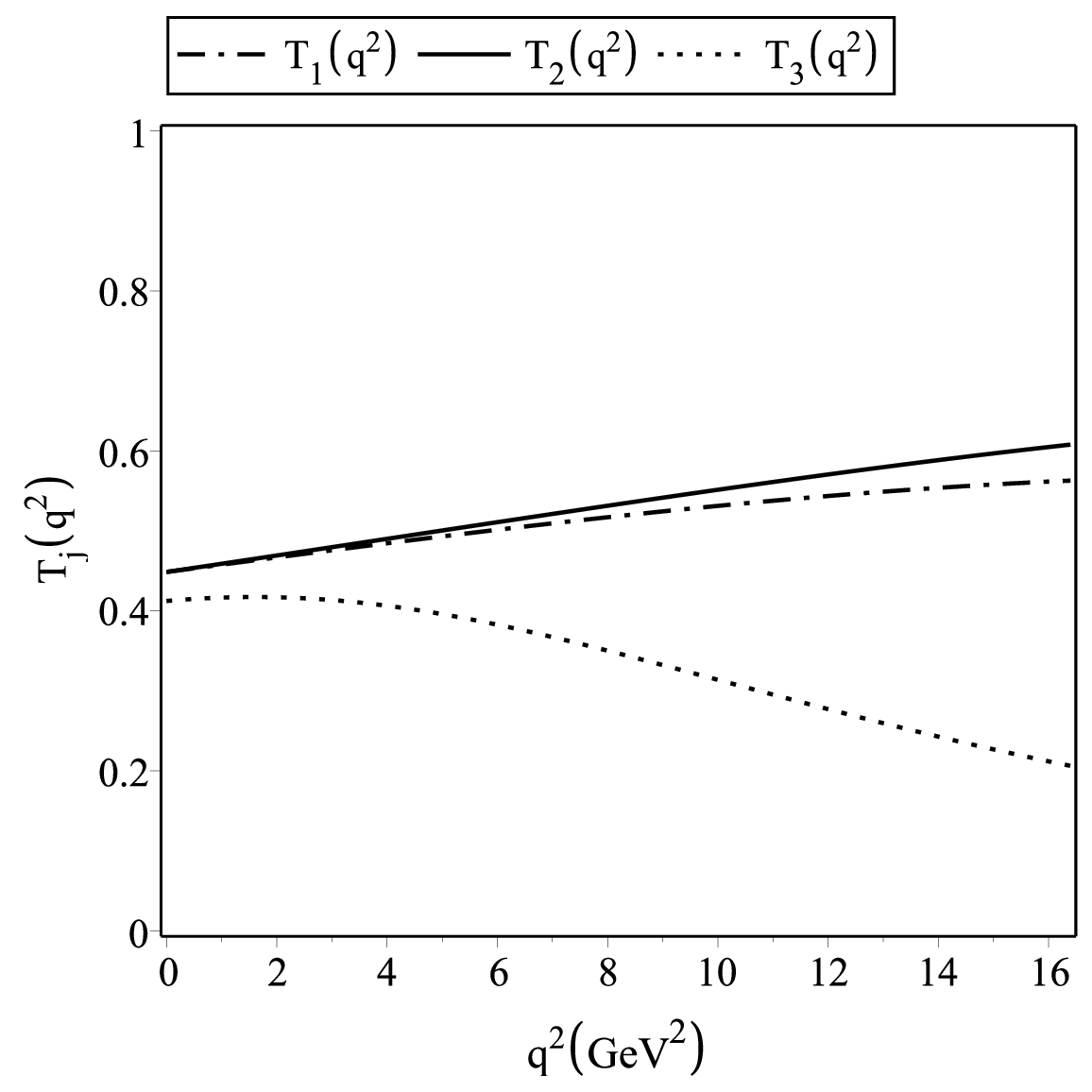}
\caption{The form factors $A, {V}_{i}$ and ${T}_{j}$ on
$q^2$.}\label{F2}
\end{figure}
Now, we can  evaluate the branching ratio values for the FCNC $B\to
a_{1} \ell^{+} \ell^{-}$ decays and the radiative $B\to a_{1}
\gamma$. For the radiative $B\to a_{1} \gamma$ transition, the
exclusive decay width is given as \cite{Safir}:
\begin{eqnarray}\label{eq42}
\Gamma(B \to a_1 \gamma)& = &  \frac{\alpha_{em}\, G_F^2 m_{b}^5}
{32 \pi^4} {|V_{tb} V_{td}^* C_{7}(m_b) T_1(0)|}^2 ~\left
(1-{m_{a_1}^2\over m_B^2}\right )^3 \left (1+ {m_{a_1}^2\over
m_B^2}\right ).
\end{eqnarray}
Also, the ratio of the exclusive--to--inclusive radiative decay
branching ratio is defined as
\begin{eqnarray}\label{eq43}
R &\equiv& {BR(B \to a_1\ \gamma)\over BR(B \to X_b\ \gamma)} =
{|T_1(0)|}^2 {\left (1- m_{a_1}^2/ m_B^2 \right )^3 \left (1+
m_{a_1}^2/ m_B^2 \right ) \over \left (1-m_d^2/ m_b^2\right )^3
\left (1+m_{d}^2/ m_b^2\right )}.
 \end{eqnarray}
$R$ is a  quantity to test the model dependence of the form factors
for the exclusive decay \cite{Safir}. Using the value of $T_{1}(0)$,
we  estimate the branching ratio $Br(B\to a_1 \gamma)=5.6\times
10^{-7}$ and the corresponding ratio $R=8.81\%$. Our prediction
means that about $8.81\%$ of the inclusive $b\to d\gamma$
branching ratio goes into $a_{1}$ channel.

For the FCNC $B\to a_{1} \ell^{+} \ell^{-}$  transition, the double
differential decay rate ${d^2\Gamma}/{dq^2 dcos\theta_{\ell}} $ is
defined in \cite{Colangelo}, where  $\theta_{\ell}$ is the angle
between the $\ell^+$  direction and the $B$ direction in the rest
frame of the lepton pair.

We  show the dependency of the differential branching ratios of
$B\to a_{1} \ell^{+} \ell^{-}\, (\ell=\mu, \tau)$ decays  on $q^2$,
with and without LD effects (see Eq. (\ref{eq23})), in Fig.
\ref{F41}.
\begin{figure}
\includegraphics[width=6cm,height=5cm]{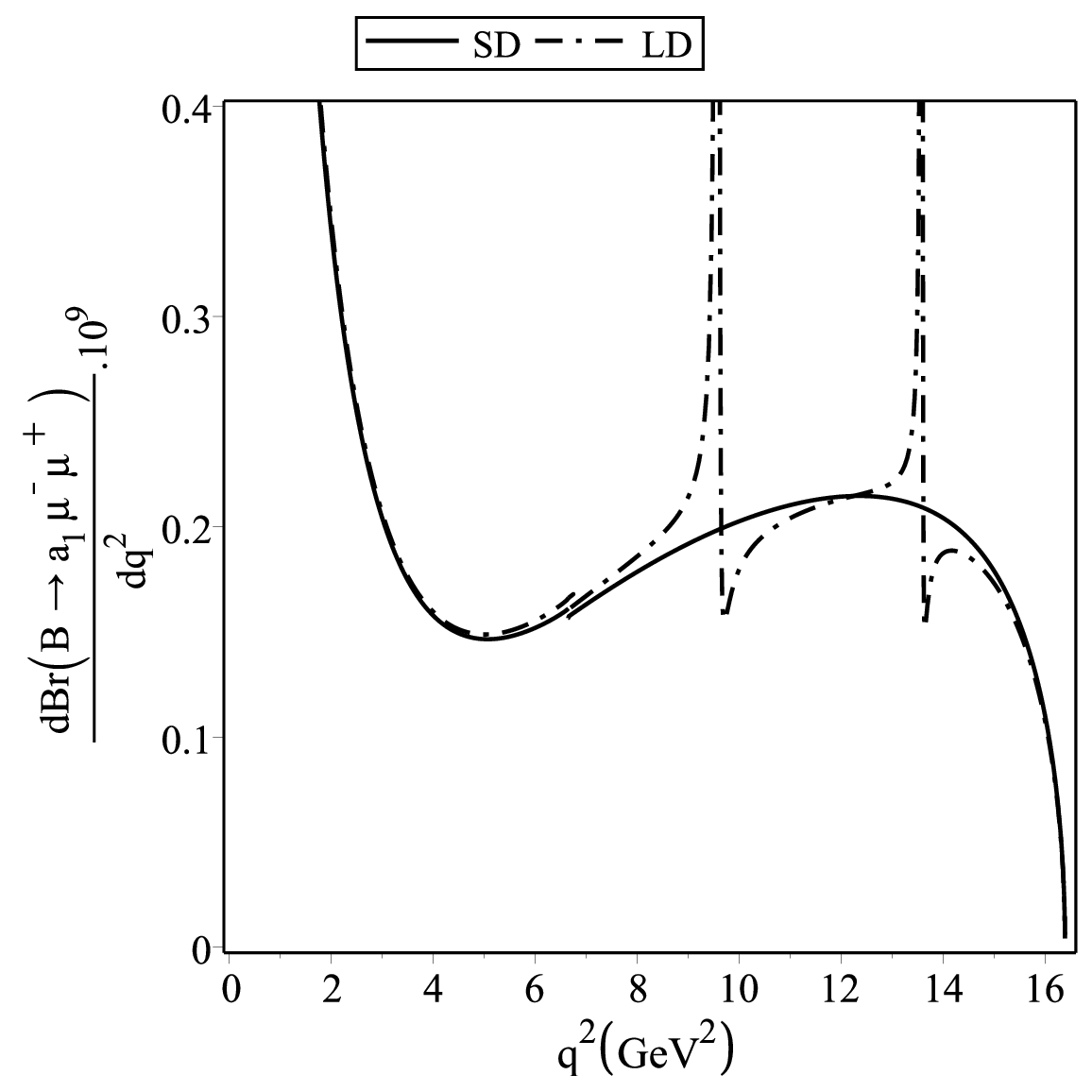}
\includegraphics[width=6cm,height=5cm]{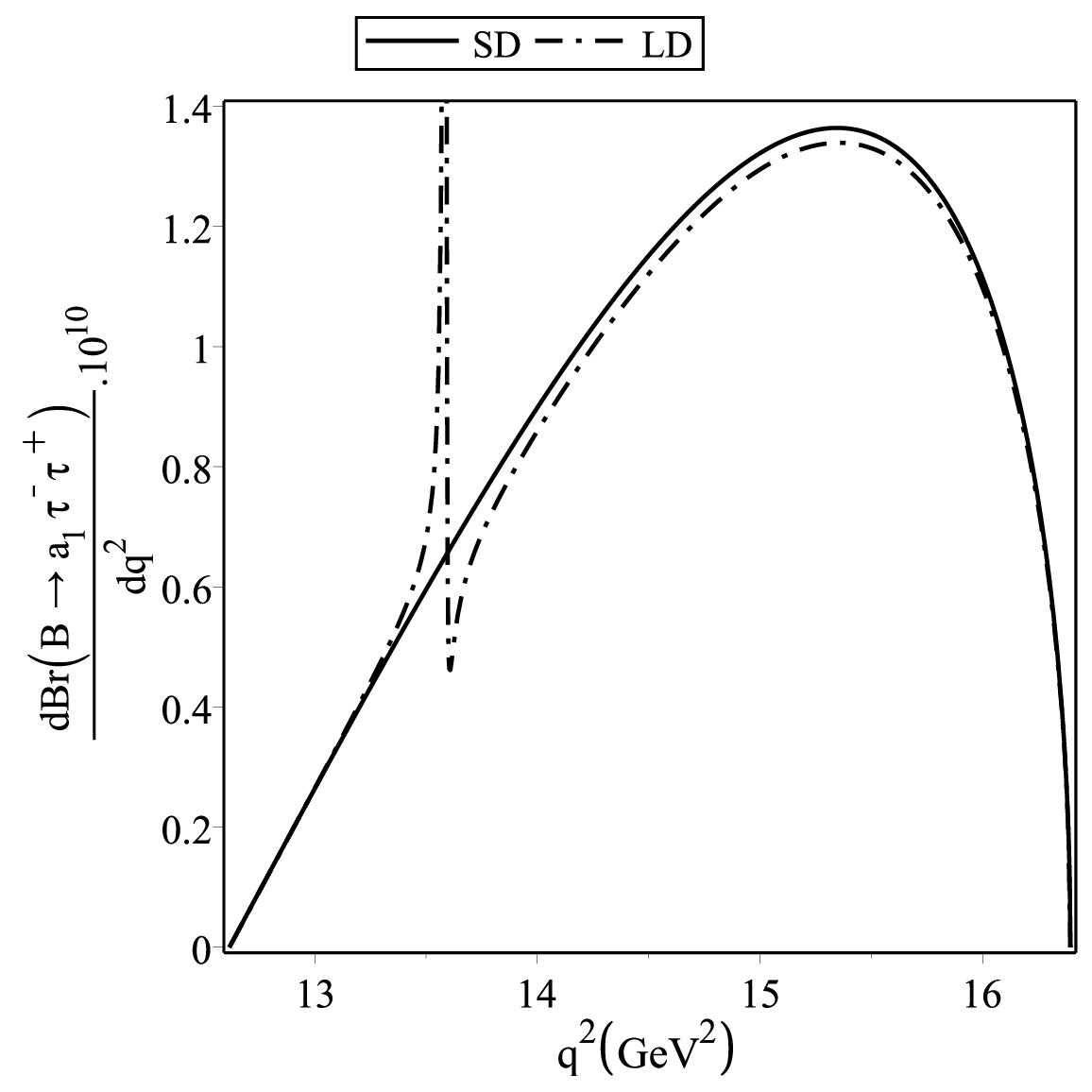}
\caption{The differential branching ratios of the semileptonic $B\to
a_1$ decays on $q^2$ with and without  LD effects. } \label{F41}
\end{figure}
In this figure, the solid and dash-dotted lines show the results
without and with the LD effects, respectively. To obtain the
branching ratio values of these  decays, some cuts around the narrow
resonances of $J/\psi$ and $\psi'$  are defined for muon as:
\begin{eqnarray}\label{eq44}
\mbox{I}: &&\ \ \ \ \ \ \ \ 2 m_{\mu} \;\leq\; \sqrt{q^2}
\;\leq\; M_{J/\psi }-0.20\,,
\nonumber\\
\mbox{II}: && M_{J/\psi}+0.04 \;\leq\; \sqrt{q^2} \;\leq\;
M_{\psi^{\prime}}-0.10\,,
\nonumber \\
\mbox{III}: &&\ \ M_{\psi^{\prime}}+0.02 \;\leq\; \sqrt{q^2}
\;\leq\; m_{B}-m_{a_1},
\end{eqnarray}
and for $\tau$, the following two regions are introduced:
\begin{eqnarray}\label{eq45}
\mbox{I}: & \ \ \ \ \ \ \ 2 m_{\tau} \;\leq\; \sqrt{q^2}
\;\leq\; M_{\psi'} - 0.02\,,
\nonumber\\
\mbox{II}: & M_{\psi'} + 0.02\; \leq\; \sqrt{ q^2}\; \leq\;
m_{B}-m_{a_1}.
\label{eqr2}
\end{eqnarray}
In Table \ref{T3},  the branching ratio values for $B\to a_{1}
\ell^{+} \ell^{-}\, (\ell=\mu, \tau)$ have been obtained using the
regions shown in Eqs. (\ref{eq44}-\ref{eq45}). The results have been
neglected for the electron since these are very close to the same as
those for the muon.
\begin{table}[th]
\caption{The branching ratios of the semileptonic $B\to a_1
\ell^+\ell^-$ decays including  LD effects in three regions.
}\label{T3}
\begin{ruledtabular}
\begin{tabular}{ccccccc}
\mbox{Mode}&I&II&III&I+II+III\\
\hline
\mbox{Br}$(B \to
a_1\mu^+\mu^-)\times10^8$&${1.92\pm0.56}$&${0.22\pm0.05}$&
${0.05\pm0.01}$&${2.19\pm0.62}$\\
$\mbox{Br}(B \to a_1\tau^+\tau^-)\times10^{9}$&$\mbox{undefined}$&
${0.10\pm0.02}$&${0.11\pm0.04}$&${0.21\pm0.06}$
\end{tabular}
\end{ruledtabular}
\end{table}

Finally, we would like to consider the  forward--backward asymmetry
$A^{FB}$ for the $B\to a_1 \ell^+\ell^- (\ell=\mu, \tau)$  decays.
The expression of the  $A^{FB}$ is given in \cite{Colangelo}. The
dependence of $A^{FB}$ for the  aforementioned  decays on $q^2$ with
and without LD effects is plotted in Fig. \ref{F42}.
\begin{figure}
\includegraphics[width=6cm,height=5cm]{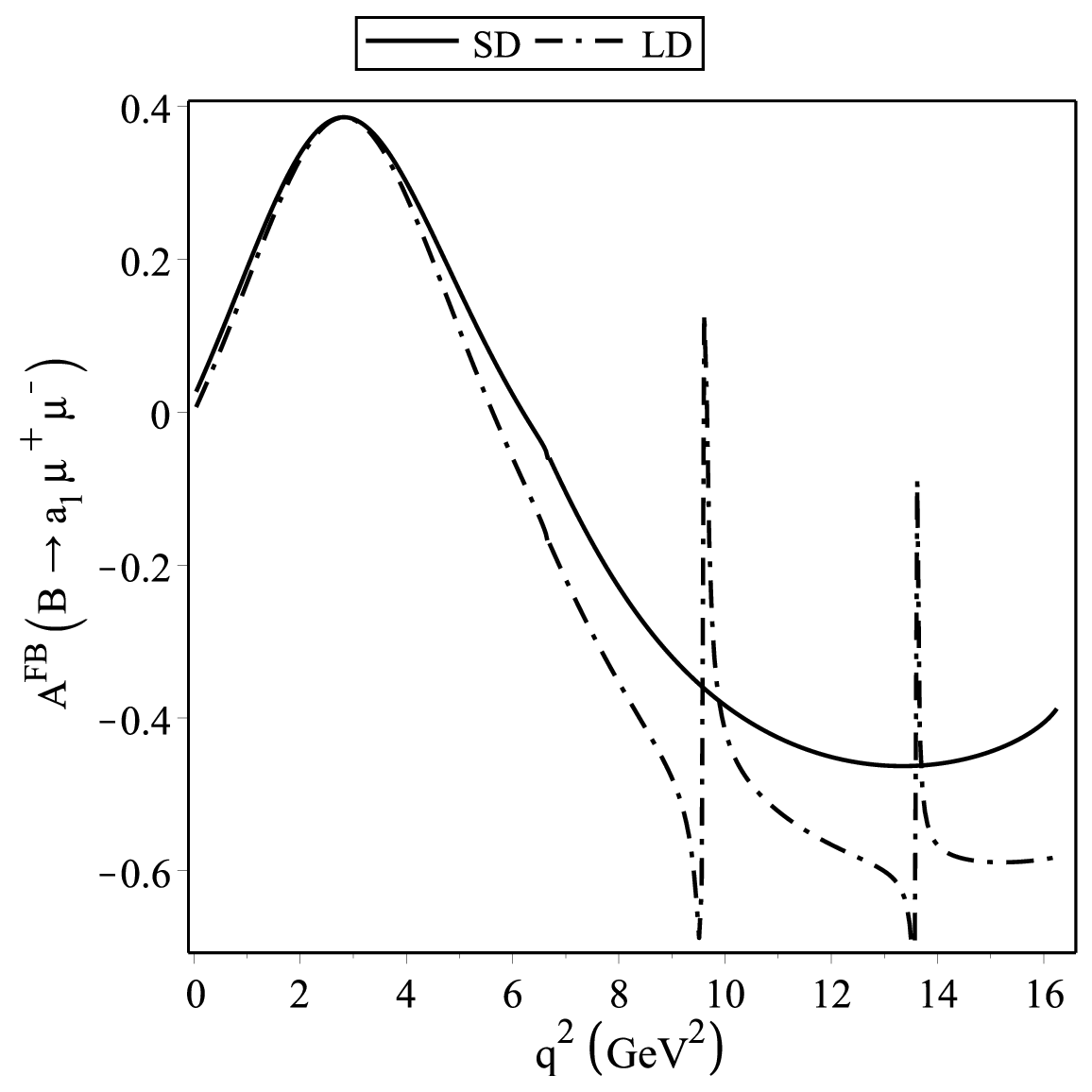}
\includegraphics[width=6cm,height=5cm]{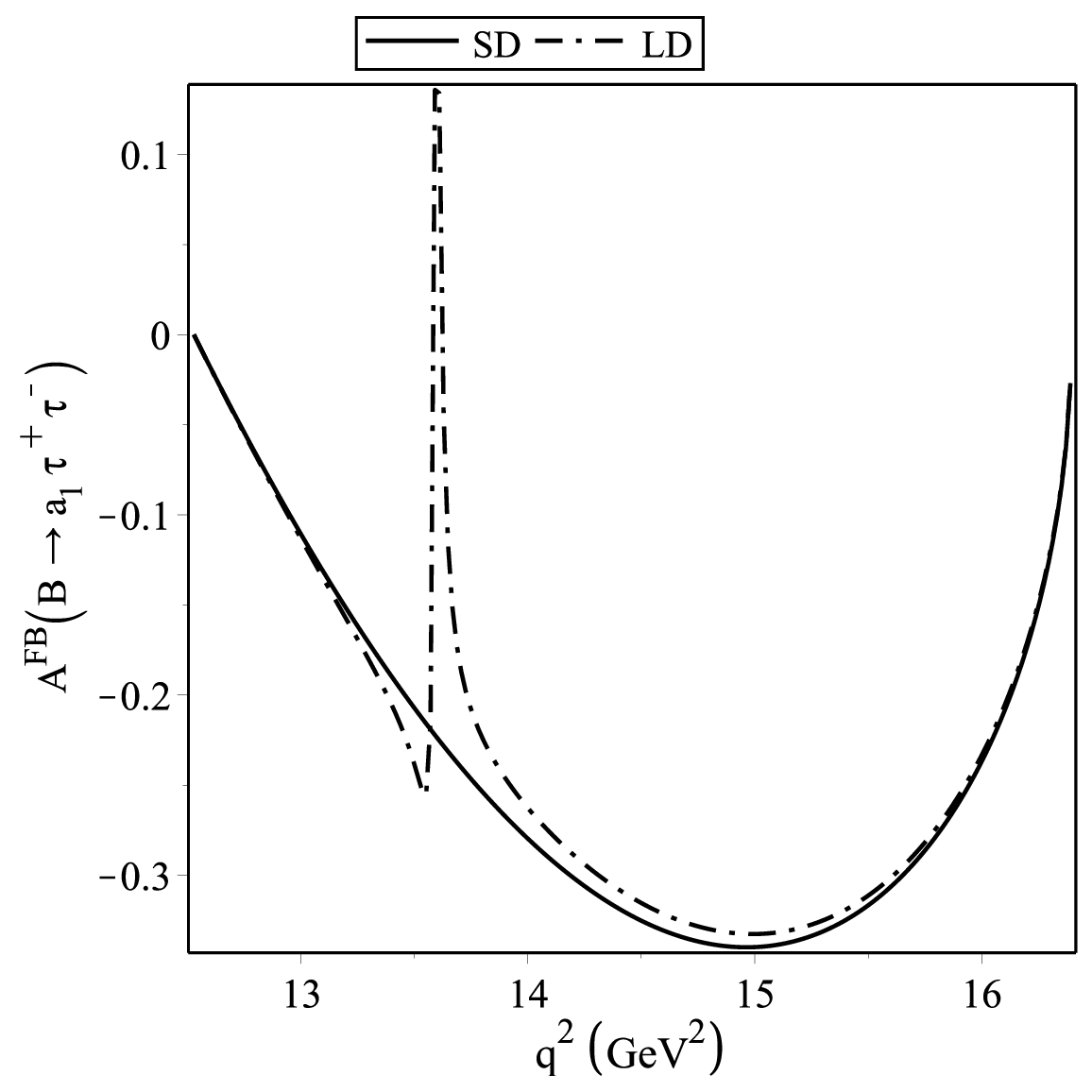}
\caption{The dependence of the forward--backward asymmetry on $q^2$.
The solid and dash-dotted lines show the results without and with
the LD effects, respectively.} \label{F42}
\end{figure}

In summary, we investigated the  form factors of the FCNC $B$ decays
into the $a_1$ axial vector meson in the LCSR approach up to the
twist--4 LCDAs. Considering both the SD and LD effects contributing
to the Wilson coefficient $ C_{9}^{\rm eff}$, we calculated the
branching ratio values for the semileptonic decays $B\to a_1 \gamma$
and $B\to a_1 \ell^{+}\ell^{-}\, (\ell=\mu, \tau)$. Finally, the
dependence of the forward--backward asymmetric of the decays $B\to
a_1 \mu^{+}\mu^{-}$ and $B\to a_1 \tau^{+}\tau^{-}$ was plotted with
respect to $q^{2}$.

\clearpage
\appendix
\section{Twist Function Definitions}\label{app:fun-def}
In this appendix, we present the definitions for $\Phi_\parallel$,
$\Phi_\perp$, $g_\perp^{(a)}$, $g_\perp^{(v)}$, $h_\parallel^{(t)}$,
$h_\parallel^{(p)}$ and $g_3$.

The  functions $\Phi_\parallel$ and  $\Phi_\perp$ for $a_1$ meson
are defined as \cite{Kwei2}:
\begin{eqnarray}
\Phi_\parallel(u) & = & 6 u \bar u \left[ a_0^\parallel + 3
a_1^\parallel\, \xi +
a_2^\parallel\, \frac{3}{2} ( 5\xi^2  - 1 ) \right],\nonumber \\
\Phi_\perp(u) & = & 6 u \bar u \left[ 1 + 3 a_1^\perp\, \xi +
a_2^\perp\, \frac{3}{2} ( 5\xi^2  - 1 ) \right],
\end{eqnarray}
where $\xi=2u-1$. Also $u$ and $\bar{u}=1-u$ refer to the momentum
fractions carried by the quark and anti-quark, respectively, in the
axial vector meson $a_1$.  These  LCDAs  are normalized as the
normalization conditions
\begin{eqnarray}\label{eqn37}
\int_0^1 du \Phi_\parallel(u) &=& 1,\nonumber\\
\int_0^1 du \Phi_\perp (u) &=& a^\perp_0.
\end{eqnarray}

Up to conformal spin $9/2$, the approximate expressions for
$g_\perp^{(a)}$, $g_\perp^{(v)}$, $h_\parallel^{(t)}$ and
$h_\parallel^{(p)}$ are taken as:
\begin{eqnarray}\label{eqn38}
g_\perp^{(a)}(u) & = &  \frac{3}{4}(1+\xi^2) +
\frac{3}{2}a_1^\parallel\, \xi^3 + \left(\frac{3}{7} \,
a_2^\parallel + 5 \zeta_{3,a_1}^V \right)\left(3\xi^2-1\right) +
\left( \frac{9}{112}a_2^\parallel + \frac{105}{16} \zeta_{3,a_1}^A
-\frac{15}{64}\zeta_{3,a_1}^V\omega_{a_1}^V
\right)  \nonumber\\
& & \times \left( 35\xi^4 - 30 \xi^2 + 3\right) + 5\Bigg[
\frac{21}{4}\zeta_{3,a_1}^V \sigma_{a_1}^V + \zeta_{3,a_1}^A
\bigg(\lambda_{a_1}^A -\frac{3}{16} \sigma_{a_1}^A\Bigg)
\Bigg]\xi(5\xi^2-3)
-\frac{9}{2} {a}_1^\perp \,\widetilde{\delta}_+ \nonumber\\
& &\times \left(\frac{3}{2}+\frac{3}{2}\xi^2+\ln u
+\ln\bar{u}\right) - \frac{9}{2} {a}_1^\perp\,\widetilde{\delta}_-\,
(3\xi + \ln\bar{u} - \ln u), \label{eq:ga}\nonumber\\
g_\perp^{(v)}(u) & = & 6 u \bar u \Bigg\{ 1 + \Bigg(a_1^\parallel +
\frac{20}{3} \zeta_{3,a_1}^A \lambda_{a_1}^A\Bigg)+
\Bigg[\frac{1}{4}a_2^\parallel + \frac{5}{3}\, \zeta^V_{3,a_1}
\left(1-\frac{3}{16}\, \omega^V_{a_1}\right)
+\frac{35}{4} \zeta^A_{3,a_1}\Bigg]  \xi (5\xi^2-1)\nonumber\\
&&+ \frac{35}{4}\Bigg(\zeta_{3,a_1}^V \sigma_{a_1}^V
-\frac{1}{28}\zeta_{3,a_1}^A \sigma_{a_1}^A \Bigg) \xi(7\xi^2-3)
\Bigg\}-18 \, a_1^\perp\widetilde{\delta}_+ \,  (3u \bar{u} +
\bar{u} \ln \bar{u} + u \ln u )\nonumber\\
& & {} - 18\, a_1^\perp\widetilde{\delta}_- \,  (u \bar u\xi +
\bar{u} \ln \bar{u} - u \ln u),\nonumber\\
h_\parallel^{(t)}(u) &= & 3a_0^\perp\xi^2+ \frac{3}{2}\,a_1^\perp
\,\xi (3 \xi^2-1) + \frac{3}{2} \Bigg[a_2^\perp \xi +
\zeta^\perp_{3,a_1}\Bigg(5-\frac{\omega_{a_1}^{\perp}}{2}\Bigg)\Bigg]\,
\xi \,(5\xi^2-3)+\frac{35}{4}\zeta^\perp_{3,a_1} \sigma^\perp_{a_1}
\nonumber\\&& \times(35\xi^4-30\xi^2+3) + 18 {a}_2^\parallel
\Bigg[\widetilde{\delta}_+ \xi -\frac{5}{8}\widetilde{\delta}_- (3\xi^2-1)\Bigg]-
\frac{3}{2}(1+ 6 {a}_2^\parallel) \Bigg( \widetilde{\delta}_+ \, \xi [2 +
\ln (\bar{u}u)]\nonumber\\
&&+\,\widetilde{\delta}_-  \, [ 1 + \xi \ln (\bar{u}/u) ]\Bigg) ,\nonumber\\
h_\parallel^{(p)}(u) & = & 6u\bar u \Bigg\{ a_0^\perp +
\Bigg[a_1^\perp +5\zeta^\perp_{3, a_1}\Bigg(1-\frac{1}{40}(7\xi^2-3)
\omega_{a_1}^{\perp} \Bigg)\Bigg] \xi+  \Bigg( \frac{1}{4}a_2^\perp
+ \frac{35}{6} \zeta^\perp_{3,a_1} \sigma^\perp_{a_1} \Bigg)
\nonumber\\ && \times (5\xi^2-1) -5{a}_2^\parallel
\Bigg[\widetilde{\delta}_+ \xi + \frac{3}{2} \widetilde{\delta}_- (1-\bar{u} u)
\Bigg]\Bigg\} - 3(1+ 6 a_2^\parallel)[\, \widetilde{\delta}_+\, (\bar{u}
\ln \bar{u} - u \ln u)\nonumber\\
& & {} + \,\widetilde{\delta}_-\,  ( u \bar{u} + \bar{u} \ln \bar{u} + u \ln
u)],
\end{eqnarray}
where
\begin{eqnarray*}
\widetilde{\delta}_\pm  ={f_{a_1}^{\perp}\over f_{a_1}}{m_{d} \pm
m_{d} \over m_{a_1}}\, ,\qquad \zeta_{3,a_1}^{V(A)} =
\frac{f^{V(A)}_{3,a_1}}{f_{a_1} m_{a_1}}.
\end{eqnarray*}
In the $SU(3)$ limit, the normalization conditions for
$g_\perp^{(a)}$, $g_\perp^{(v)}$, $h_\parallel^{(t)}$ and
$h_\parallel^{(p)}$ are defined as:
\begin{eqnarray}
\int_0^1 du g_\perp^{(a)}(u)     &=&\int_0^1 dug_\perp^{(v)}(u)=1\,,
\nonumber\\
\int_0^1 du h_\parallel^{(t)}(u) &=& a^\perp_0,
\nonumber\\
\int_0^1 du h_\parallel^{(p)}(u) &=& a^\perp_0 + \widetilde{\delta}_-.
\end{eqnarray}

The definition of the function $g_{3}(u)$  is as follows
\cite{Ball}:
\begin{eqnarray}
g_{3}(u)=6u(1-u)+(1-3\xi^2)\left[\frac{1}{7}{a}_2^\parallel-\frac{20}{3}\frac{f^{A}_{3,a_1}}{f_{a_1}m_{a_1}}\right].
\end{eqnarray}

\clearpage
\section{Form Factor Expressions }\label{app:form factors}

In this appendix, the explicit expressions for the form factors of
the FCNC $B\to a_1$ decays are presented.
\begin{eqnarray*}
A(q^{2})&=& \frac{f_{a_1}
m_{b}}{4\,m_{B}^{2}\,f_{B}}\,(m_{a_1}-m_{B})\Bigg\{\frac{f_{a_1}^\perp}{f_{a_1}}
\int_{u_0}^{1} du~\frac{9\,\Phi_\perp (u)}{u}e^{s(u)}
+\frac{m_{b}}{4m_{a_1}}\int_{u_0}^{1}du~\frac{{g_\perp^{(v)\prime}(u)}}{u}e^{s(u)}\nonumber\\
&-&\frac{m_{b}}{4m_{a_1}}\int_{u_0}^{1}du~\frac{{g_\perp^{(v)}(u)}}{u^2}\left[1+\frac{\delta_{1}(u)-8m_{a_1}^{2}}{M^{2}}\right]e^{s(u)}
+\frac{f_{a_1}^\perp m_{a_1}^{2}}{f_{a_1}}\int_{u_0}^{1}du~\frac{32~\bar{h}{_\parallel^{(t)(ii)}(u)}}{M^2}e^{s(u)}\Bigg\},\nonumber\\
V_{1}(q^{2})&=&
-\frac{m_{b}}{8\,m_{B}^{2}\,f_{B}}\,\frac{f_{a_1}^\perp}{(m_{B}-m_{a_1})}\Bigg\{\frac{1}{2}\int_{u_0}^{1}du
~\frac{7\,\Phi_\perp
(u)\,\delta_{1}(u)}{u}e^{s(u)}+2\,m_{a_1}^{2}\int_{u_0}^{1}du\frac{h_\parallel^{(p)}(u)}{u}
~e^{s(u)}\nonumber\\
&-&3\frac{f_{a_1}}{f_{a_1}^\perp}m_{a_1}m_{b}\int_{u_0}^{1}du~\frac{{g_\perp^{(a)}(u)}}{u}~e^{s(u)}-4\,m_{a_1}^{3}m_{b}
\int_{u_0}^{1}du~\frac{\bar{g}{_3}^{(ii)}(u)}{u^{2}M^{2}}e^{s(u)}
-8\,m_{a_1}^{2}\nonumber \\
&\times& \int_{u_0}^{1}du~
\frac{\bar{h}{_\parallel^{(t)(ii)}}(u)}{u^{2}}~e^{s(u)}
+4m_{a_1}^{2}\frac{m_{b}}{f_{a_1}^\perp}\int_{u_0}^{1}du\int {\cal
D}\,\underline{\alpha}\left[\frac{f_{3, a_1}^A\,{\cal
A}(\alpha_{i})-
f_{3, a_1}^V\,{\cal V}(\alpha_{i})}{\kappa^{2}M^{2}}\right]e^{s(\kappa)}\Bigg\},\nonumber \\
V_{2}(q^{2})&=& -\frac{f_{a_1}^\perp
m_{b}}{4\,m_{B}^{2}\,f_{B}}\,{(m_{B}-m_{a_1})}\Bigg\{18\int_{u_0}^{1}du
\frac{\Phi_\perp (u)}{u}e^{s(u)}+\frac{4f_{a_1}m_{a_1}m_{B}}{f_{a_1}^\perp}\int_{u_0}^{1}du~\frac{\phi_{a}(u)}{u^2\,M^2} e^{s(u)} \nonumber\\
&+& 4m_{a_1}^{2}
\int_{u_0}^{1}du\frac{{h_\parallel^{(p)}(u)}}{u}(1+2u)e^{s(u)}
-8f_{a_1}\,m_{a_1}^{3}m_{b}\int_{u_0}^{1}du\frac{\bar{g}{_3}^{(ii)}(u)}{u^{3}M^{4}}~e^{s(u)} \nonumber\\
&+&\frac{16f_{a_1}}{f_{a_1}^\perp}  m_{a_1}m_{b}
 \int_{u_0}^{1}du~\frac{{\Phi_{\|}}^{(i)}(u)}{u^{2}M^{2}}~e^{s(u)}
-16m_{a_1}^{2}\int_{u_0}^{1}du
\frac{\bar{h}{_\parallel^{(t)(ii)}}(u)}{u^{2}}\left[\frac{2\,\delta_{3}(u)}{u\,M^4}-\frac{3}{2\,M^2}
\right.\nonumber\\&+&\left. \frac{\delta_{1}(u)}{4u\,M^4}\right]e^{s(u)}\Bigg\},\nonumber\\
V_{0}(q^{2})&=&V_{3}(q^{2})+\frac{m_{b}}{8\,m_{B}^{2}\,f_{B}}\,\frac{f_{a_1}^\perp
q^2}{m_{a_1}}\Bigg\{9\int_{u_0}^{1}du \frac{\Phi_\perp
(u)}{u}~e^{s(u)}+2f_{a_1}\,m_{a_1}m_{B}\int_{u_0}^{1}du
~\frac{\phi_{a}(u)}{u^2\,M^2}
~e^{s(u)}\nonumber\\
&-&4 m_{a_1}^{2}
\int_{u_0}^{1}du~\frac{{h_\parallel^{(p)}(u)}}{u}~(1-u)~e^{s(u)}
-4\frac{f_{a_1}}{f_{a_1}^\perp}\,m_{a_1}^{3}m_{b}\int_{u_0}^{1}du~\frac{\bar{g}{_3}^{(ii)}(u)}{u^{3}M^{4}}~(1-u)~e^{s(u)}\nonumber\\
&+&\frac{16f_{a_1}}{f_{a_1}^\perp}m_{a_1}m_{b}\int_{u_0}^{1}du~\frac{{\Phi_{\|}^{(i)}}(u)}{u^{2}M^{2}}~e^{s(u)}
+8m_{a_1}^{2}\int_{u_0}^{1}du~
\frac{\bar{h}_{\parallel}^{(t)(ii)}(u)}{u^{2}}\left[\frac{2\,\delta_{3}(u)}{u\,M^4}
-\frac{1}{M^2}+(1-u)\right.\nonumber\\
&\times& \left. \left(-\frac{1}{M^2}
+\frac{\delta_{1}(u)}{2u\,M^4}\right)\right]~e^{s(u)}\Bigg\},\nonumber\\
T_{1}(q^{2})&=&-\frac{f_{a_1}
m_{b}}{8\,m_{B}^{2}\,f_{B}}\Bigg\{m_{b}(\frac{f_{a_1}^\perp}{f_{a_1}}+8)\int_{u_0}^{1}du
\frac{\Phi_\perp (u)}{u}~e^{s(u)}
-3m_{a_1}\int_{u_0}^{1}du~g_{\perp}^{(a)}(u)~e^{s(u)}\nonumber\\
&+&4m_{a_1}
\int_{u_0}^{1}du~\frac{\phi_{a}(u)}{u}~e^{s(u)}-\frac{f_{a_1}}{
m_{a_1}}
\int_{u_0}^{1}du~\frac{{g_\perp^{(v)\prime}(u)\,\delta_{5}(u)}}{u}~e^{s(u)}
-\frac{m_{a_1}}{8}\int_{u_0}^{1}du~\frac{{g_\perp^{(v)\prime}(u)}}{u}\nonumber\\
&\times &\left[7-\frac{\delta_{5}(u)(8u-1)}{M^2}+
\frac{u\delta_{2}(u)-\delta_{4}(u)}{2\,m_{a}^2}\right] ~e^{s(u)}
+4m_{a_1}\int_{u_0}^{1}du\frac{{\Phi_{\|}^{(i)}}(u)}{u}~e^{s(u)}
-4m_{a_1}^{3}\nonumber\\
&\times&
\int_{u_0}^{1}du~\frac{\bar{g}{_3}^{(ii)}(u)}{u\,M^{2}}~e^{s(u)}
-\frac{16f_{a_1}^\perp}{f_{a_1}}m_{a_1}^{2}\,m_b\int_{u_0}^{1}du
\frac{\bar{h}{_\parallel^{(t)(ii)}}(u)}{u\,M^2}~e^{s(u)}+4\frac{f_{3,
a_1}^A}{f_{a_1}}\int_{u_0}^{1}du\int
{\cal D}\underline{\alpha}\nonumber\\
&\times& \frac{u\,{\cal A}(\alpha_{i})}{\kappa^2}
\left[1+\frac{\delta_{1}(\kappa)}{M^2}\right] ~e^{s(\kappa)}\Bigg\},\nonumber\\
T_{2}(q^{2})&=&\frac{m_{b}}{m_{B}^{2}\,f_{B}}\frac{f_{a_1}}{m_{a_1}^2-m_{B}^2}\Bigg\{\frac{m_{b}f_{a_1}^\perp}
{f_{a_1}}\int_{u_0}^{1}du ~\frac{\Phi_\perp
(u)\delta_{1}(u)}{u}~e^{s(u)}
+\frac{1}{2}m_{a_1}\int_{u_0}^{1}du \frac{~g_{\perp}^{(a)}(u)}{u}\nonumber\\
&\times&
\left[\delta_{1}(u)+4\,\delta_{5}(u)\right]~e^{s(u)}-\frac{1}{16}
m_{a_1}
\int_{u_0}^{1}du~\frac{{g_\perp^{(v)\prime}(u)\,\delta_{2}(u)}}{u}~e^{s(u)}
+\frac{1}{2}{m_{a_1}}\int_{u_0}^{1}du~\frac{\phi_{a}(u)}{u}\nonumber\\
&\times&
\delta_{1}(u)e^{s(u)}+{m_{a_1}}\int_{u_0}^{1}du~\frac{g_\perp^{(v)}(u)}{u^2}\left[\delta_{6}(u)+
\frac{\delta_{1}(u)\delta_{5}(u)}{M^2}+\frac{u \delta_{2}(u)}{2}
\Bigg(1+\frac{\delta_{3}(u)
}{u\,M^2}+\frac{\delta_{7}(u)}{u}\Bigg) \right.\nonumber\\
 &+& \left.u
\Bigg(m_{a_1}^2-2\delta_{1}(u)+\frac{\delta_{4}(u)}{2}+\frac{\delta_{5}(u)\delta_{1}(u)}{u\,M^2}\Bigg)\right]e^{s(u)}
+2m_{a_1}^{3}\int_{u_0}^{1}du~\frac{\bar{g}{_3}^{(ii)}(u)}{u^2}\left[5-\frac{\delta_{5}(u)}{M^2}\right]e^{s(u)}\nonumber\\
&-&2m_{a_1}\int_{u_0}^{1}du~\frac{{\Phi_{\|}^{(i)}(u)\delta_{2}(u)}}{u}e^{s(u)}
-8\frac{f_{a_1}^\perp}{f_{a_1}}\,m_{a_1}^{2}\,m_b\int_{u_0}^{1}du~\frac{\bar{h}{_\parallel^{(t)(ii)}}(u)}{u^2}
\left[1+\frac{\delta_{2}(u)}{M^2}\right]e^{s(u)}\nonumber\\
&-&\frac{1}{2}\frac{f_{3, a_1}^V}{f_{a_1}} \int_{u_0}^{1}du\int
{\cal D}\,\underline{\alpha}~\frac{{\cal V}(\alpha_{i})}{\kappa^2}
\left[\delta_{4}(\kappa)+\frac{\delta_{1}(\kappa)\,\delta_{2}(\kappa)}{M^2}
+u m_{a_1}^2\left(1+\frac{\delta_{2}(\kappa)}{M^2}\right)\right]
~e^{s(\kappa)}\nonumber\\
&-&\frac{f_{3, a_1}^A}{f_{a_1}} \int_{u_0}^{1}du\int {\cal
D}\,\underline{\alpha}~\frac{{\cal A}(\alpha_{i})}{\kappa^2}
\left[m_{a_1}^2\frac{\delta_{2}(\kappa)}{M^2}-4\,u\left(m_{a_1}^2+\frac{\delta_{4}(\kappa)}{4}
+\frac{\delta_{1}(\kappa)\delta_{2}(\kappa)}{M^2}\right)\right]e^{s(\kappa)}\Bigg\},\nonumber\\
T_{3}(q^{2}) &=& -\frac{f_{a_1} m_{b}}{4\,m_{B}^{2}\,f_{B}}
\Bigg\{\frac{8f_{a_1}^\perp}{f_{a_1}}m_{b}\int_{u_0}^{1}du
~\frac{\Phi_\perp (u)}{u}~e^{s(u)} -4m_{a_1}\int_{u_0}^{1}du
\frac{~g_{\perp}^{(a)}(u)}{u}~e^{s(u)}
+2m_{a_1}^3\nonumber\\
&\times&
\int_{u_0}^{1}du~\frac{\bar{g}{_3}^{(ii)}(u)}{u^2}\left[\frac{8}{M^2}-\frac{2\,\delta_{5}(u)}{M^4}\right]e^{s(u)}
-\frac{1}{4\, m_{a_1}}
\int_{u_0}^{1}du~\frac{{g_\perp^{(v)\prime}(u)}}{u}~\left[\frac{7}{2}\delta_{2}(u)+m_{a_1}^2 \right.\nonumber\\
&-&\left. \frac{\delta_{1}(u)}{4\,u} \right]e^{s(u)}
-m_{a_1}\int_{u_0}^{1}du~\frac{\phi_{a}(u)}{u^2}\left[\frac{1}{u}+\frac{u\,\delta_{1}(u)+2\,\delta_{2}(u)}{M^2}\right]e^{s(u)}-
4m_{a_1}\int_{u_0}^{1}du~\frac{{\Phi_{\|}}^{(i)}(u)}{u^2}\nonumber\\
&\times&\left[\frac{u\,\delta_{5}(u)-\delta_{2}(u)}{M^2}
-1\right]e^{s(u)}-\frac{1}{4m_{a_1}}\int_{u_0}^{1}du\frac{{g_\perp^{(v)}(u)}}{u}
\left[\frac{3\,\delta_{1}(u)}{u\,M^2}-\frac{\delta_{1}(u)}{m_{a_1}^2}+
\frac{5\,\delta_{5}(u)-7\,\delta_{3}(u)}{M^2}\right.\nonumber\\
&+&\left.\frac{\delta_{2}(u)}{M^2}-\frac{\delta_{1}(u)^{2}}
{m_{a_1}^2\,M^2}\right]e^{s(u)}
+\frac{16f_{a_1}^\perp}{f_{a_1}}\,m_{a_1}^{2}\,m_b\int_{u_0}^{1}du~
\frac{\bar{h}{_\parallel^{(t)(ii)}}(u)}{u^2\,M^2}\left[8+\frac{2}{u}+\frac{\delta_{2}(u)}{u\,M^2}\right]e^{s(u)}\Bigg\},
\end{eqnarray*}
where
\begin{eqnarray*}
u_{0} &=&\frac{1}{2m_{a_1}^2} \left[\sqrt{(s_0-m_{a_1}^2-q^2)^2 +4 m_{a_1}^2 (m_b^2-q^2)} -\left(s_0-m_{a_1}^2-q^2\right)\right],\nonumber\\
s(u)&=&-\frac{1}{uM^2}\left[m_b^2+u\,\bar{u}m_{a_1}^2-\bar{u}q^2\right]+\frac{m_B^2}{M^2},\nonumber\\
\delta_{1}(u)&=& m_{a_1}^2(u+2)+\frac{m_{b}^2}{u}+\frac{q^2}{u}\nonumber,\\
\delta_{2}(u)&=&u\,m_{a_1}^2-\frac{m_{b}^2}{u}+q^2\, \frac{u-\bar{u}}{u},\nonumber\\
\delta_{3}(u)&=& \frac{m_{b}^2}{u}-2q^2\, \frac{\bar{u}}{u},\nonumber\\
\delta_{4}(u)&=& 2\,m_{a_1}^2(u+1)+2q^2,\nonumber\\
\delta_{5}(u)&=&u\,m_{a_1}^2-\frac{m_{b}^2}{u}+\frac{q^2(u-2)}{u},\nonumber\\
\delta_{6}(u)&=& 2\,m_{a_1}^2(u+1)+q^2\frac{\bar{u}}{u},\nonumber\\
\delta_{7}(u)&=&-2\frac{m_{b}^2}{u}+\frac{q^2}{u},\nonumber\\
{f}^{(i)}(u)&\equiv&\int_0^u f(v) dv,\nonumber\\
{f}^{(ii)}(u)&\equiv&\int_0^u dv\int_0^v d\omega f(\omega),\nonumber\\
\phi_a&=& \int_0^u \left[\Phi_\parallel - g_\perp^{(a)} (v)\right]dv,\nonumber\\
\kappa&=&\alpha_1+u\alpha_3.
\end{eqnarray*}

\end{document}